\documentclass[onecolumn]{aastex631}

\usepackage{amsmath,amssymb}
\usepackage{graphicx}
\usepackage{booktabs}
\usepackage{longtable}
\usepackage{array}
\usepackage{multirow}
\usepackage{url}
\usepackage{hyperref}

\newcommand{\kms}{\,km\,s$^{-1}$}

\newcommand{\Vadj}{\ensuremath{V_{\rm adj}}}
\newcommand{\Vbary}{\ensuremath{V_{\rm bary}}}

\newcommand{\omrad}{\ensuremath{\omega}}

\newcommand{\HI}{H\,{\sc i}}
\newcommand{\corpusdoi}{10.5281/zenodo.23503221}

\begin{document}

\title{A Unified \HI\ Rotation Curve Database for 122
Dwarf and Irregular Galaxies}

\author{D.~C.~Flynn}
\affiliation{EPS Research, Laurel, MD 20708, USA}
\email{davidflynn@eps-research.com}
\correspondingauthor{D. C. Flynn}

\received{}
\revised{}
\accepted{}
\submitjournal{PASP}

\begin{abstract}
We present a unified \HI\ rotation curve database for 122 dwarf and irregular
galaxies drawn from three Local Volume surveys, the Local Volume \HI\ Survey
(LVHIS; 32 galaxies), VLA-ANGST (29) and LITTLE THINGS (26), and from the
WALLABY DR2 kinematic catalogue (35).
The database provides standardised kinematic parameters, distance estimates,
morphological classifications, and rotation curve data in machine-readable
JSON, JSONL, and CSV formats with a documented schema, supporting
retrieval-augmented generation (RAG) applications and cross-survey kinematic
analysis. Quality tiers distinguish 26 galaxies with published tilted-ring
rotation curves and per-point uncertainties from 96 automated-pipeline,
single-ring or profile-width entries. Three worked examples demonstrate corpus
queries, including evaluation of the $\omega$ estimator for DDO~154
(LITTLE THINGS). This work is presented as a data resource; no new dynamical
model is proposed. This version describes corpus v1.3, which corrects the
LITTLE THINGS curves (v1.1), the WALLABY entries (v1.2) and the LVHIS
entries and reference (v1.3) of v1.0. The database and all
computation scripts are available at Zenodo (\doi{\corpusdoi}).
\end{abstract}


% ══════════════════════════════════════════════════════════════════════════
\section{Introduction}
\label{sec:intro}
% ══════════════════════════════════════════════════════════════════════════

Dwarf and irregular galaxies occupy a unique position in galactic
kinematic studies. Their shallow potential wells, gas-dominated
baryonic inventories, and high dark matter fractions
\citep[e.g.,][]{Oh2015,Koribalski2018} make them sensitive probes of
the relationship between baryonic mass and observed kinematics — a
relationship that remains contested in the context of modified gravity
frameworks \citep{Milgrom1983,McGaugh2016} and $\Lambda$CDM halo models
\citep{NFW1997}.

The \HI\ rotation curves of nearby dwarfs have been
measured by several major interferometric surveys. The Local Volume \HI\
Survey \citep[LVHIS;][]{Koribalski2018} used the Australia Telescope
Compact Array (ATCA) to observe 82 galaxies within 10\,Mpc. The
VLA-ANGST survey \citep[VLA-ANGST;][]{Ott2012} targeted 35 dwarf
galaxies in the Local Volume with the Very Large Array (VLA). The
LITTLE THINGS survey \citep[Local Irregulars That Trace Luminosity
Extremes;][]{Hunter2012} imaged 41 nearby dwarf irregular galaxies with
the VLA, and \citet{Oh2015} derived rotation curves and mass models for 26
of them. The WALLABY pilot survey with ASKAP \citep{Koribalski2020,Westmeier2022}
extends coverage to larger distances with a blind survey approach; its
kinematic catalogue \citep{Deg2022,Murugeshan2024} provides automated
tilted-ring models.

Despite the scientific importance of this population, no unified
kinematic corpus covering all four surveys in a machine-readable,
schema-consistent format has been published. Individual survey catalogs
use different column conventions, distance methods, and quality flags,
making cross-survey comparison and retrieval-augmented generation (RAG)
applications challenging.

The present work addresses this gap. We assemble and standardise
rotation curve data from all four surveys into a unified JSON/CSV
corpus with a documented schema, quality tiers, and omega-readiness
flags. The $\omega$ correction is treated throughout as an empirical kinematic descriptor; no modification to Newtonian or relativistic gravity is proposed or implied.

This work forms part of the following series:

\begin{enumerate}
\item Flynn \& Cannaliato (2025) introduced the \omrad\ kinematic
  correction and demonstrated its empirical performance across 84 SPARC
  Q=1 disk galaxies \citep{Flynn2025}.

\item Flynn (2026a) examined the \omrad\ correction across the same
  84 SPARC galaxies with full baryonic decomposition
  \citep{Flynn2026paper2}.

\item Flynn (2027) published a unified galaxy rotation curve corpus
  of 438 galaxies from SPARC, THINGS, LITTLE THINGS, and WALLABY DR2
  in \textit{Astronomy and Computing} \citep{Flynn2026corpus}; its
  corrected release v7.3 \citep{Flynn2026v73} is the source of the
  WALLABY entries used here.

\item This work provides the dwarf/irregular extension corpus (122
  galaxies) and flags the 24 LITTLE THINGS galaxies that are eligible
  for the \omrad\ estimator.
\end{enumerate}

Beyond its value as a data resource, the corpus provides the data needed
to examine whether the kinematic regularity described by \omrad\ in
massive spirals carries over to dark-matter-dominated dwarfs, where
baryonic disk physics is different.

% ══════════════════════════════════════════════════════════════════════════
\section{Survey Sources and Selection}
\label{sec:surveys}
% ══════════════════════════════════════════════════════════════════════════

\subsection{LVHIS}
\label{sec:lvhis}

The Local Volume \HI\ Survey \citep{Koribalski2018} provides ATCA \HI\
observations of 82 galaxies within approximately 10\,Mpc. Table~9 of
\citet{Koribalski2018} lists 47 tilted-ring kinematic solutions for 38 of
them, either from the LVHIS kinematic models or compiled from the literature.
We include every galaxy with a Table~9 solution except six massive spirals
and ellipticals with peak rotation velocities of 150\kms\ or more and
non-Irr/Im/dIrr morphological classifications: NGC\,253
(SABc, 200\kms), NGC\,1313 (SBd, 220\kms), NGC\,4945 (SBcd, 174\kms),
NGC\,5128/Cen\,A (S0, 260\kms), NGC\,5236/M\,83 (Sc, 150\kms), and
Circinus (Sb, 161\kms). This leaves 32 galaxies; the selection is applied
in the deposited build script (\texttt{build\_dwarf\_v13.py}) from a verbatim
transcription of Table~9 that is included in the deposit.

Each LVHIS entry is a single-point seed summary of one Table~9 solution: the
rotation velocity near the maximum fitted radius $R_{\rm max}$, with the
fitted inclination, position angle and dynamical mass. For the nine galaxies
with more than one Table~9 solution, the summary uses the solution with the
largest $R_{\rm max}$; this is a fixed catalogue convention, not a choice of
the preferred or best-quality fit. Where Table~9 gives a range (for example,
the position angle of a warped disc), the summary value is the midpoint,
flagged in the record. All Table~9 solutions are stored verbatim in each
record (\texttt{table9\_solutions}), and the source of the selected solution
is recorded (\texttt{kinematic\_source}); 16 of the 32 selected solutions are
literature fits compiled in Table~9. All LVHIS entries are assigned quality
tier~2.

\subsection{VLA-ANGST}
\label{sec:angst}

The VLA-ANGST survey \citep{Ott2012} targeted 35 dwarf galaxies in the
Local Volume with the VLA in B and C configurations. We ingest all 29
galaxies with published kinematic parameters. Each VLA-ANGST entry is a
single-point profile-width estimate from the catalogue \HI\ line width
$W_{20}$ and inclination $i$,
\begin{equation}
V_{\rm rot} = \frac{\max(W_{20} - 16\,\mathrm{km\,s^{-1}},\ 0)}{2\sin i},
\label{eq:vla}
\end{equation}
where the 16\kms\ term corrects for instrumental broadening and turbulence;
all VLA-ANGST entries are assigned quality tier~2. The VLA-ANGST records carry
no measured \HI\ radius (\texttt{rhi\_kpc} is null). Their radial fields are
distance-based placeholders, $R_{\rm seed} = \max[0.5\,(D/\mathrm{Mpc}),\ 1]$\,kpc
for \texttt{r\_max\_kpc} (and the single data point) and $0.1\,R_{\rm seed}$ for
\texttt{r\_min\_kpc}, and must not be used as measured \HI\ radii or as
spatially resolved rotation-curve points.

\subsection{LITTLE THINGS}
\label{sec:lt}

\citet{Oh2015} derived rotation curves for 26 of the 41 LITTLE THINGS
galaxies \citep{Hunter2012} from 2D tilted-ring fits to the bulk velocity
fields with the GIPSY task ROTCUR, corrected for asymmetric drift. All 26
are included. The rotation curves are the observed (Type\,=\,Data) rows of
the published table (VizieR J/AJ/149/180; 877 points in total) with their
published uncertainties; the pseudo-isothermal model curves of
\citet{Oh2015} are stored separately (\texttt{model\_iso\_oh2015}) in the 25
records for which model rows are published (DDO\,43 has none) and are not
part of the rotation curve data. Radii are sorted monotonically on ingest. All 26
LITTLE THINGS galaxies are assigned quality tier~1.

\subsection{WALLABY DR2}
\label{sec:wallaby}

The WALLABY DR2 kinematic catalogue \citep{Deg2022,Murugeshan2024}
provides rotation curves from the WALLABY Kinematic Analysis Proto-Pipeline
(WKAPP) for galaxies detected in the ASKAP pilot fields
\citep{Koribalski2020,Westmeier2022}. The WALLABY entries are taken from
the corrected unified corpus v7.3 \citep{Flynn2026v73}, which carries the
published WKAPP optimised models with the catalogue's per-ring velocity
uncertainties, inclinations, position angles and systemic velocities from
the same catalogue row, and distances \texttt{dist\_h} (Hubble flow,
$H_0 = 70$\kms\,Mpc$^{-1}$) from the DR2 source catalogue.
A WALLABY galaxy is included if it meets any one of three criteria:
peak rotation velocity below 80\kms, mean rotation velocity below
60\kms, or distance below 10\,Mpc. Applied to the 203 WALLABY galaxies of
v7.3, this selects 35. The selection is by velocity or distance, not by
volume: 33 of the 35 lie beyond 10\,Mpc (range 8.8--78.3\,Mpc), and three
lie beyond 50\,Mpc. WALLABY entries are automated pipeline products and are
assigned quality tier~2. The inner rings are affected by beam smearing at
the 30\,arcsec ASKAP resolution.
The 35 WALLABY entries in this corpus do not include HI masses,
dynamical masses, HI radii or optical Hubble types
(\texttt{mhi\_log\_msun}, \texttt{log\_mdyn}, \texttt{mhi\_mdyn\_ratio},
\texttt{rhi\_kpc}, \texttt{hubble\_type}). These are omissions from this
corpus; the WALLABY source catalogue does provide HI masses
\citep{Murugeshan2024}. Each record states the omission, and flags beam
smearing, Hubble-flow distances, velocity-derived morphological classes and
the absence of baryonic decomposition, in its \texttt{known\_issues} field.
The \texttt{morph\_class} of a
WALLABY entry is estimated from its peak rotation velocity.

% ══════════════════════════════════════════════════════════════════════════
\section{Corpus Construction}
\label{sec:construction}
% ══════════════════════════════════════════════════════════════════════════

\subsection{Schema Design}
\label{sec:schema}

Each galaxy entry in the corpus is a JSON object with standardised
fields covering identification, coordinates, morphology, distance,
kinematics, HI properties, rotation curve data, provenance and quality
metadata; a 32-column flat CSV summarises each galaxy.
Table~\ref{tab:schema} lists the core fields and their units.
The full schema is documented in the Zenodo deposit
(\doi{\corpusdoi}).

\begin{table}[ht]
\centering
\caption{Core corpus schema fields.\label{tab:schema}}
\begin{tabular}{lll}
\hline\hline
Field & Type & Description \\
\hline
galaxy & string & Identifier \\
survey & string & LVHIS $|$ VLA\_ANGST $|$ LITTLE\_THINGS $|$ WALLABY \\
distance\_mpc & float & Distance [Mpc] \\
distance\_method & string & TRGB $|$ Cepheid $|$ TF $|$ membership $|$ Hubble \\
inc\_deg & float & Inclination [$^\circ$] \\
pa\_deg & float & Position angle [$^\circ$] \\
vrot\_max\_kms & float & Peak rotation velocity [km\,s$^{-1}$] \\
n\_points & int & Rotation curve points \\
r\_min\_kpc & float & Innermost radius [kpc] \\
r\_max\_kpc & float & Outermost radius [kpc] \\
rhi\_kpc & float & HI radius [kpc] \\
quality\_tier & int & 1 = published RC with per-point errors; 2 = pipeline/seed/estimate \\
omega\_ready & bool & Eligible for the \omrad\ estimator \\
data & array & \textit{Rad} (kpc), \textit{Vrot} (km\,s$^{-1}$), and \textit{errV} (LITTLE THINGS) \\
     &       & or \textit{e\_Vrot}, \textit{e\_Vrot\_avg}, \textit{e\_Vrot\_inc} (WALLABY) \\
\hline
\end{tabular}
\end{table}

\subsection{Quality Tiers}
\label{sec:tiers}

We define two quality tiers:

\textbf{Tier~1} (26 galaxies): published tilted-ring rotation curves with
per-point uncertainties. All LITTLE THINGS galaxies.

\textbf{Tier~2} (96 galaxies): automated pipeline products, single-ring
seed values, or profile-width estimates. The 35 WALLABY entries
(multi-ring WKAPP models), the 32 LVHIS entries (single-point seed summaries
of Table~9 solutions of \citealt{Koribalski2018}; Section~\ref{sec:lvhis})
and the 29 VLA-ANGST entries (profile-width estimates from \citealt{Ott2012}).

\subsection{Omega-Readiness}
\label{sec:omega_ready}

A galaxy is flagged \texttt{omega\_ready=True} if it satisfies:
(1) $n_{\rm points} \geq 5$; (2) quality tier~1; and (3) the rotation
field is classified as kinematically regular. This yields 24 omega-ready
galaxies, all from LITTLE THINGS. The flag marks eligibility for the
estimator only; it does not imply that \omrad\ has been validated for
these galaxies.

Two LITTLE THINGS galaxies are explicitly excluded from the omega-ready
subsample despite meeting the point-count criterion:

\textbf{IC\,10}: The rotation curve extends only to $R_{\rm max} =
0.54$\,kpc, too small for a reliable boundary fit; the second ring
($R = 0.080$\,kpc) has a negative rotation velocity, indicating
disturbed kinematics; and the innermost point ($V_1 = 1.95$\kms,
$\sigma_{V_1} = 12.71$\kms) is unconstrained.

\textbf{NGC\,3738}: The inclination of $22.6^\circ$ falls below the
reliable threshold for inclination-corrected rotation velocities, and the
maximum rotation velocity ($132.7$\kms) is anomalously high for a
dIrr classification.

In three omega-ready galaxies (DDO\,87, DDO\,216 and WLM) the innermost
velocity is also smaller than its uncertainty. For these the inner term of
the estimator contributes less than 5\% of the outer term, so the low
signal-to-noise of the innermost point has little effect on \omrad.

% ══════════════════════════════════════════════════════════════════════════

\section{Worked Examples}
\label{sec:examples}

To demonstrate corpus utility, we provide three representative query examples.
Full worked examples with retrievable JSON contexts are provided in
\texttt{rag\_examples\_v1.json} in the Zenodo deposit.

\subsection{Example 1: Single-Galaxy Kinematic Query}
\label{sec:ex1}

\textit{Query:} What is the inclination, position angle, and peak rotation
velocity of IC\,5152, and what group does it belong to?

\textit{Result:} IC\,5152 (LVHIS) is a dwarf irregular (Im) at a TRGB
distance of $D = 1.97$\,Mpc and a member of the Sculptor group. Table~9 of
\citet{Koribalski2018} lists two solutions for this warped disc; the corpus
summary uses the one with the larger fitted radius, from \citet{Kirby2012}:
inclination $49^\circ$, position angle $234.5^\circ$ (the midpoint of the
published range $271$--$198^\circ$), and rotation velocity
$V_{\rm rot} = 59.5$\kms\ near the maximum fitted radius of 4.0\,kpc. Both
published solutions are returned in the record's \texttt{table9\_solutions}
field; the midpoint is a summary and does not describe the warped geometry.

\subsection{Example 2: Population Filter}
\label{sec:ex2}

\textit{Query:} Which galaxies in the Centaurus\,A group have peak rotation
velocities below 80\kms?

\textit{Result:} Querying \texttt{group\_member = "Cen\_A"} and
\texttt{vrot\_max\_kms < 80} returns five galaxies, all from the LVHIS
component (ESO\,174-G001, ESO\,325-G011, ESO\,381-G020, NGC\,5237 and
UKS\,1424-460), with Hubble types IBm, Im? or I0, distances 3.4--5.4\,Mpc and
rotation velocities 22--75\kms. These are single-point Table~9 seed
values, not rotation curves.

\subsection{Example 3: The \omrad\ Estimator Applied to DDO\,154}
\label{sec:ex3}

\textit{Query:} Apply the Flynn \& Cannaliato (2025) two-boundary
estimator to DDO\,154 (LITTLE THINGS).

\begin{figure}[!h]
\centering
\includegraphics[width=0.95\columnwidth]{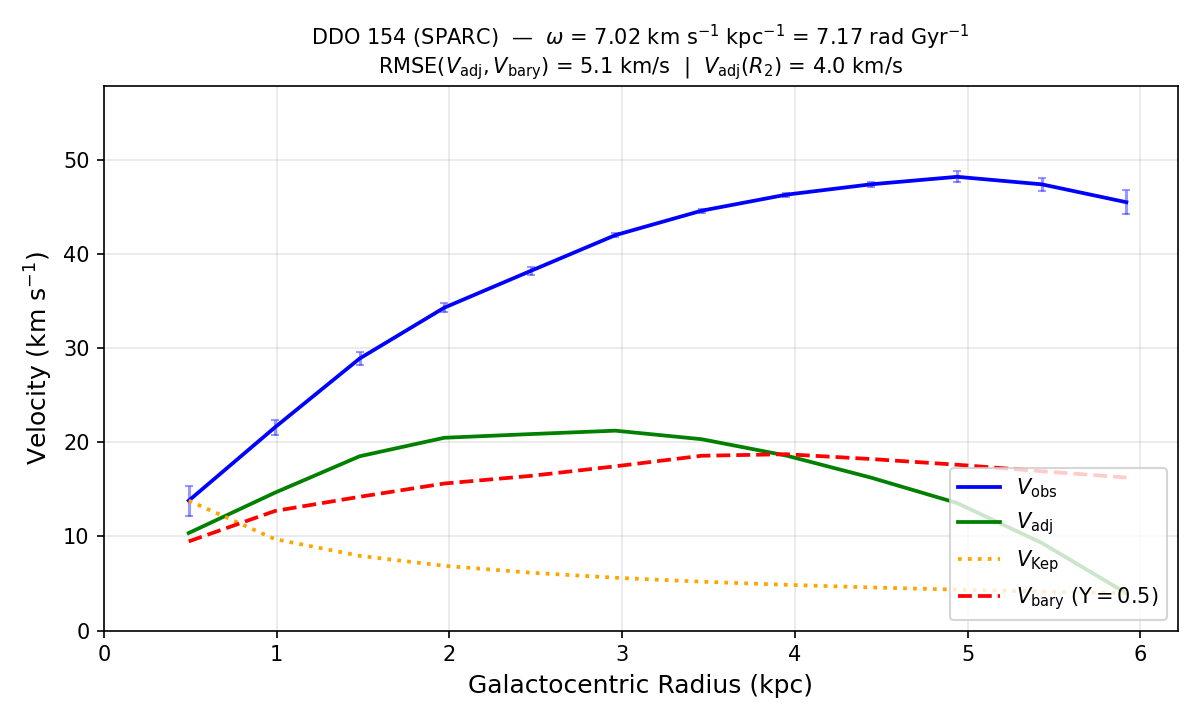}
\caption{Rotation curve decomposition for DDO\,154 from the SPARC
rotmod file \citep{Lelli2016}. Blue: $V_{\rm obs}$; green:
$V_{\rm adj} = V_{\rm obs} - R\omega$; orange dotted: $V_{\rm Kep}$;
red dashed: $V_{\rm bary}$ ($\Upsilon = 0.5$).
The SPARC rotmod boundary points ($R_1 = 0.49$\,kpc, $R_2 = 5.92$\,kpc)
give $\omega = 7.015$\,km\,s$^{-1}$\,kpc$^{-1}$ $= 7.174$\,rad\,Gyr$^{-1}$,
compared with 6.440\,rad\,Gyr$^{-1}$ from the LITTLE THINGS boundary points
(Section~\ref{sec:ex3}); the two curves come from independent kinematic
analyses with different radial extents. RMSE$(V_{\rm adj}, V_{\rm bary})
= 5.11$\kms.
\label{fig:ddo154}}
\end{figure}

\textit{Result:} Using the innermost and outermost observed boundary
points of the LITTLE THINGS curve in this corpus
($R_1 = 0.140$\,kpc, $V_1 = 6.20$\kms; $R_2 = 7.890$\,kpc,
$V_2 = 50.51$\kms), the angular velocity offset is:

\begin{equation}
\omega = \frac{V_2}{R_2} - \frac{V_1}{R_1}\left(\frac{R_1}{R_2}\right)^{3/2}
= \frac{50.51}{7.890} - \frac{6.20}{0.140}\left(\frac{0.140}{7.890}\right)^{3/2}
= 6.297\ \mathrm{km\,s^{-1}\,kpc^{-1}} = 6.440\ \mathrm{rad\,Gyr^{-1}},
\label{eq:omega_ex}
\end{equation}

using $1\ \mathrm{km\,s^{-1}\,kpc^{-1}} = 1.0227\ \mathrm{rad\,Gyr^{-1}}$.
This is a per-galaxy value; no comparison with the SPARC population is made
here. Figure~\ref{fig:ddo154} shows the four-curve decomposition
for DDO\,154 using the SPARC rotmod file \citep{Lelli2016},
which provides the baryonic decomposition that the LITTLE THINGS records
lack.

\section{Discussion}
\label{sec:discussion}
% ══════════════════════════════════════════════════════════════════════════

We discuss the omega results and corpus limitations below.

\subsection{Omega: Per-Galaxy Values}
\label{sec:discussion_omega}

Per-galaxy \omrad\ values for the 24 omega-ready LITTLE THINGS galaxies,
in km\,s$^{-1}$\,kpc$^{-1}$ and rad\,Gyr$^{-1}$, are provided in
\texttt{omega\_results\_dwarfs.csv} in the Zenodo deposit. All 24 values are
positive. They are eligibility values computed from the observed boundary
points; no population statistic, outer-gap diagnostic, or comparison with
the SPARC sample is reported, pending independent validation of the
boundary-ring convention for dwarf rotation curves.

This corpus provides the data needed to test whether $\omega$ correlates
with baryonic surface density, dark matter fraction, or other physical
parameters across the dwarf-to-spiral transition. Such tests require
full baryonic decomposition of the LITTLE THINGS galaxies, which is
not available in the present corpus but is a natural target for future
cross-matching with available HI mass models. Cosmological simulations
such as RAMSES \citep{Teyssier2002} could provide an independent test
of whether a quantity like $\omega$ arises from disk formation physics.

\subsection{Limitations}
\label{sec:limitations}

The present corpus has several limitations that future versions should
address:

\textbf{No baryonic decomposition for LITTLE THINGS.} The omega-ready
subsample computes \Vadj\ but cannot complete the full RMSE$(\Vadj,
\Vbary)$ comparison without Vgas/Vdisk baryonic decomposition. The
DDO\,154 example uses the SPARC rotmod file as an exception. Future
versions should cross-match LITTLE THINGS with available HI mass models
to enable full baryonic validation.

\textbf{LVHIS and VLA-ANGST in seed mode.} The 61 LVHIS and VLA-ANGST
galaxies carry single-point velocity estimates rather than full rotation
curves. The LVHIS inclinations and position angles are summary values of one
Table~9 solution (midpoints where a range is published) and should not
replace radial geometry in deprojection or dynamical modelling. FITS-mode ingestion of the published data cubes would yield
multi-point profiles for many of these galaxies, significantly increasing
the omega-ready subsample.

\textbf{Missing vsys for LVHIS.} Systemic velocities are absent for
all 32 LVHIS entries in the current corpus. These can be ingested from
\citet{Koribalski2018} in a future version.

\textbf{Distance range of the WALLABY subset.} The WALLABY entries are
selected by velocity or distance, not volume: 33 of 35 lie beyond 10\,Mpc
and three beyond 50\,Mpc. The corpus as a whole is therefore not a
volume-limited Local Volume sample; users can apply a cut on
\texttt{distance\_mpc}.

\textbf{WALLABY beam smearing.} The WALLABY curves are automated
pipeline products at 30\,arcsec resolution; their inner rings are affected
by beam smearing, and morphological classes are estimated from velocity.

% ══════════════════════════════════════════════════════════════════════════
\section{Summary}
\label{sec:summary}
% ══════════════════════════════════════════════════════════════════════════

We have presented the Dwarf/Irregular Galaxy \HI\ Rotation Curve Corpus
v1.3, a unified kinematic database of 122 dwarf and irregular galaxies
assembled from LVHIS, VLA-ANGST, LITTLE THINGS, and WALLABY DR2. The
principal results are:

\begin{enumerate}

\item The corpus provides 122 galaxies in unified JSON/CSV/JSONL format
  with a documented schema, quality tiers, and omega-readiness flags.
  All entries pass schema validation.

\item 24 LITTLE THINGS galaxies are classified omega-ready (quality
  tier~1, $n_{\rm points} \geq 5$, regular rotation fields).

\item Three worked examples demonstrate corpus queries: single-galaxy
  kinematic lookup (IC\,5152), population filter (Cen\,A group dwarfs),
  and the \omrad\ estimator applied to DDO\,154 (LITTLE THINGS),
  giving $\omega = 6.297$\,km\,s$^{-1}$\,kpc$^{-1}$
  $= 6.440$\,rad\,Gyr$^{-1}$.

\end{enumerate}

The corpus and all scripts are available at Zenodo
(\doi{\corpusdoi}; all versions: \doi{10.5281/zenodo.20320361}).

\textbf{Changes from version 1 of this preprint.} Version 1 described
corpus v1.0. Corpus v1.1 removed 839 pseudo-isothermal model rows of
\citet{Oh2015} that had been interleaved with the 877 observed LITTLE THINGS
rows, and corpus v1.2 replaced the WALLABY entries with the published WKAPP
models via the corrected unified corpus v7.3 and re-applied the selection
(41 to 35 WALLABY galaxies; 129 to 123 in total). Corpus v1.3 rebuilt the
LVHIS entries from Table~9 of \citet{Koribalski2018} under the declared rules
of Section~\ref{sec:lvhis}, replacing seed values that in nine cases combined
or averaged two published solutions, and removed NGC\,5408, which has no
Table~9 solution (122 in total); the LVHIS reference was previously given
incorrectly. Examples~1 and~2 change accordingly: IC\,5152 now reports one
published solution, and in the Cen\,A filter NGC\,5408 is replaced by
ESO\,174-G001. This version also corrects
the LITTLE THINGS method (ROTCUR, not 2DBAT), the km\,s$^{-1}$\,kpc$^{-1}$ to
rad\,Gyr$^{-1}$ conversion (multiply by 1.0227), and the worked-example
values; and it withdraws the v1 population statistics (median $\omega$),
the outer-gap comparison and the SPARC comparison, which used model-curve
boundary points and an inverted unit conversion.

% ══════════════════════════════════════════════════════════════════════════
\begin{acknowledgments}
The SPARC database is maintained by F.~Lelli and S.~McGaugh (CWRU).
This work was conducted as independent research by EPS Research and
received no external funding.

\textit{AI assistance.} The numerical results, tables, and figures
were produced using author-run Python workflows on a local HPC cluster
(Node1: Intel i9-14900K, 128\,GB DDR5, Ubuntu 24.04), directly from
published survey data, including AI-assisted scripts as described below.
Large language models were used for manuscript review,
\LaTeX\ formatting, literature cross-checking, and language editing;
for drafting the corpus v1.2 and v1.3 build and verification scripts, which
the author ran and checked on Node1; and for independent audits of the v1.2
and v1.3 releases and of this manuscript, whose findings the author reviewed
and acted on.
All scientific content, interpretations, and conclusions are the sole
responsibility of the author.
\end{acknowledgments}

% ══════════════════════════════════════════════════════════════════════════
\software{
Python \citep{python3},
NumPy \citep{numpy},
SciPy \citep{scipy},
Matplotlib \citep{matplotlib},
Astropy \citep{astropy2022}
}

% ══════════════════════════════════════════════════════════════════════════

\end{document}